\documentclass[12pt]{article}
\usepackage{sc3conf}
\usepackage{latexsym}
\usepackage{amssymb}
\usepackage{graphicx}
\usepackage{times}

\newcommand{\be}{\begin{equation}}
\newcommand{\ee}{\end{equation}}
\newcommand{\bea}{\begin{eqnarray}}
\newcommand{\eea}{\end{eqnarray}}
\newcommand{\nn}{\nonumber}

\newcommand{\intsum}{\sum\!\!\!\!\!\!\!\int}

  \let\g=\gamma \let\d=\delta

\let\t=\tau 
\let\ph=\varphi

\newcommand{\halb}{\frac{1}{2}}
\newcommand{\hal}{{\textstyle\frac{1}{2}}}

\newcommand{\vf}{\varphi}
\newcommand{\ve}{\epsilon}
\begin{document}
\raggedbottom
\title{Quantum corrections to the mass and central charge of solitons\\ in 
1+1 dimensions}

\authors{A.~S.~Goldhaber,\adref{1}
A.~Rebhan,\adref{2}\break P.~van~Nieuwenhuizen,\adref{1} R.~Wimmer\adref{2}}

\addresses{\1ad ~C.~N.~Yang Institute for Theoretical
Physics, SUNY at Stony Brook, NY 11794
, USA
  \nextaddress \2ad ~Institut f\"ur Theoretische Physik, Technische
Universit\"at Wien, 
A-1040 Vienna, Austria}

\maketitle

\begin{abstract}
We first discuss how the longstanding confusion in the literature concerning 
one-loop quantum corrections to 1+1 dimensional 
solitons has finally been resolved.
Then we use 't Hooft and Veltman's dimensional regularization to compute the 
kink mass,
and find that chiral domain wall fermions, 
induced by fermionic zero modes, lead to 
spontaneous parity violation and an anomalous 
contribution to the central charge
such that the BPS bound becomes saturated. 
On the other hand, Siegel's dimensional 
reduction
shifts this anomaly to the counter terms 
in the renormalized current multiplet. 
The $\gamma\cdot j$ superconformal anomaly is located in 
an evanescent counter term, and imposing 
supersymmetry, this counter term induces 
the same anomalous contribution to the 
central charge. Next we discuss  a new 
regularization scheme: local mode regularization. The local energy
density computed in this scheme satisfies the BPS equality 
(it is equal to the local
central charge density). In an appendix we give a very detailed account of the 
DHN method to compute soliton masses applied to the supersymmetric kink.
\end{abstract}

\section{Introduction}
Quantum corrections to solitons were
of great interest in the
1970's and 1980's 
\cite{Dashen:1974cj,Gervais:1976zg,Faddeev:1978rm}, 
and again in the last few years, due to
the present  activity in quantum field theories with
dualities between extended objects and pointlike objects.  
Dashen, Hasslacher, and Neveu
\cite{Dashen:1974cj}, in a 1974 article that has become a classic, 
computed the one-loop corrections
 to the mass of the bosonic kink in
$\phi^4$ field theory and to the bosonic soliton in
sine-Gordon theory.  For the latter, there exist exact
analytical methods associated with the complete
integrability of the system, authenticating the perturbative
calculation.  Our work here uses general principles but
focuses on the kink, for which exact results are not
available.  Dashen et al. put the object (classical
background field corresponding to kink or to sine-Gordon soliton) in
a box of length
$L$ to discretize the
 continuous  spectrum, and used mode number regularization
(equal numbers of modes in the topological and trivial
sectors, including the zero mode in this counting) for the
ultraviolet divergences.  They imposed periodic boundary
conditions (PBC) on the meson field which describes the
fluctuations around the trivial or topological vacuum
solutions, and added a logarithmically divergent mass
counter term whose finite part
 was fixed by requiring absence of tadpoles in the trivial
background. They found for the one-loop correction to the kink mass 
\begin{equation} \Delta M^{(1)} = 
\sum \frac{1}{2} \hbar \omega_n - \sum \frac{1}{2}
\hbar
\omega_n^{(0)} +
\delta M = - \hbar m \left( \frac{3}{2 \pi} -
\frac{\sqrt{3}}{12}
\right) <0
\label{mass}  \ \ ,
\end{equation} where $m$ is the mass of the meson in the
trivial background and $\delta M$
the counterterm induced by renormalizing $m$. 
This result remains unchallenged. 

The supersymmetric (susy) case, as well as the general case
including fermions, proved more difficult. The action reads 
\begin{equation}\label{Lss}
 {\cal L}=- \frac{1}{2} \left(
\partial_{\mu}\varphi
\right)^{2}-  \frac{1}{2}
\bar{\psi}\!\!\not\!\partial\psi -
\frac{1}{2} U^{2} - c\frac{1}{2}
\frac{dU}{d\phi}\bar{\psi}\psi
\ \ ,\end{equation} where $- \frac{1}{2} U^{2}=
-\frac{\lambda}{4}(\varphi^{2}-\mu^{2}/\lambda)^{2}
$,  the meson mass is 
$m=\mu\sqrt{2}$, and $c=1$ for supersymmetry. Dashen et al.
did not publish the fermionic corrections to the
soliton mass, stating ``The actual computation of [the
contribution to]
$M^{(1)}$ [due to fermions] can be carried out along the
lines of the Appendix. As the result is rather complicated
and not particularly illuminating we will not give it here''
(page 4137 of
\cite{Dashen:1974cj}).

Several authors have since performed the calculation of $M^{(1)}$
for the susy kink, and found different answers. It became clear
that the answers depended on the choice of boundary conditions
(BC) for the fluctuation fields, more precisely on the BC for
the fermions. Moreover, it also became clear that one obtained
different answers if one used different regularization schemes.
At present these issues are believed to be fully understood
as follows.

{\it Boundary conditions:}
Boundary conditions distort fields near the boundary. This distortion
creates spurious boundary energy which should be subtracted from
the total energy in order to obtain the true mass of the kink.
There are several ways to avoid the spurious boundary energy
\begin{itemize}
\item[(i)]one may first compute the energy density ${\mathcal E}(x)$
and then integrate over a region which contains the kink but
stays away from the boundaries
\cite{Shifman:1998zy}; 
\item[(ii)]one may average overs {\em sets} of BC such that
in the average the boundary energy cancels \cite{Goldhaber:2000ab}. 
One such set of BC
for fermions which has been studied in detail consists of
periodic BC ($\psi^\pm(-L/2)=\psi^\pm(L/2)$),
antiperiodic BC ($\psi^\pm(-L/2)=-\psi^\pm(L/2)$),
twisted periodic BC ($\psi^\pm(-L/2)=\psi^\mp(L/2)$), and
twisted antiperiodic BC ($\psi^\pm(-L/2)=-\psi^\mp(L/2)$);%
\footnote{Strictly speaking, these BC should be called
even and odd rather than periodic and antiperiodic.}
\item[(iii)]one may choose a set of BC which have no boundary.
By this cryptic statement we mean BC which put the system
on a circle (more precisely a M\"obius strip) such that
the system becomes translationally invariant and one cannot
identify a point where the boundary is present \cite{Nastase:1998sy}.
In principle such BC could still lead to delocalized (homogeneously
spread out) boundary energy, but this does not occur \cite{Goldhaber:2002mx}.
By using the $Z_2$ symmetry
$\varphi_K(-x)=-\varphi_K(x)$ of the kink background, one such set of
BC has been identified to be the twisted (anti)periodic BC
in the kink sector;
\item[(iv)]one may first consider a background which contains
both a kink ($\mathrm K$) and an antikink ($\overline{\mathrm K}$)
with periodic BC, and then divide the answer for the mass
of this compound ${\mathrm K}\overline{\mathrm K}$ system by 2
\cite{Schonfeld:1979hg,Goldhaber:2000ab}. (Putting
a kink next to an antikink, there is a small cusp
in the background where the kink is joined to the antikink,
but for large distances the effect of this cusp can be neglected.
One can also find an exact solution which is everywhere smooth
and has periodic BC (a ``sphaleron'') but this involves
transcendental functions.) In fact, if one begins with periodic
BC for the fermions in this ${\mathrm K}\overline{\mathrm K}$ system, one
finds that the mode solutions have either twisted periodic
or twisted antiperiodic BC in between.
\end{itemize}

{\it Regularization schemes:}
Several well-known regularization schemes have been applied to the
calculation of the quantum kink mass and the quantum central
charge. To regulate the various sums over zero-point energies
one has used: mode number cutoff, energy-momentum cutoff,
heat-kernel techniques, $\zeta$-function techniques,
{}'t Hooft-Veltman's dimensional regularization,
Siegel's dimensional regularization (``dimensional reduction'').
To regulate Feynman graphs, one has used higher-space-derivative
regularization with factors $(1-\partial_x^2/M^2)$
(this regularization of the kinetic terms but not the interactions
preserves susy, although it breaks Lorentz invariance\footnote{Because
the anticommutator of two supersymmetry charges never produces
a Lorentz generator, it is possible to preserve supersymmetry
while breaking Lorentz symmetry.})
and again dimensional regularization.
It has turned out that the reason some of these schemes give 
incorrect answers is that they were applied incorrectly: one
naively applied the rules which had been developed for a trivial
background to the kink background. After proper modification,
these schemes all now yield the same answers. It is of some
interest (and useful for avoiding errors in future calculations)
to point out the required modifications
of all these schemes. In the following,
however, we concentrate on discussing in detail the
two variants of dimensional regularization
as well as a newly proposed method to study
the local energy distribution of the quantum mass,
local mode regularization \cite{Goldhaber:2001rp,Wimmer:2001yn}.

The {}'t Hooft-Veltman dimensional regularization
can be employed
in a susy preserving manner by embedding the minimally susy kink
in $1\le d\le2$ spatial dimensions. This leads to new physics, namely
spontaneous parity breaking and chiral domain wall fermions,
which provide a new explanation \cite{Rebhan:2002yw}
for the origin of the anomalous
contribution \cite{Shifman:1998zy} to the central charge of the susy kink. 
Siegel's
dimension reduction, on the other hand, where $d\le1$, obtains this anomaly
from an evanescent counter term to the superconformal current,
which gives rise to an anomalous nonconservation at the quantum
level of the conformal version of the central-charge current
\cite{Rebhan:2002yw}.

\section{Dimensional regularization and reduction}

\subsection{One-loop bosonic kink mass}

Probably the most elegant regularization scheme to avoid the
difficulties of mode regularization in a finite box and
the possibility of boundary energy is dimensional regularization
by embedding the 1+1 dimensional kink in $n=d+1$ dimensions
as a domain wall.

As has been shown in Ref.~\cite{Parnachev:2000fz}, this
reproduces correctly the one-loop quantum 
mass of the bosonic 1+1 dimensional kink,
as well as the surface tension of the higher-dimensional
kink domain walls \cite{Rebhan:2002uk}.

By analytic continuation of the number of extra transverse
dimensions ($d-1$) of a kink domain wall, no further regularization is
needed. Denoting the momenta pertaining to the extra dimensions
by $\ell$ and reserving $k$ for the momentum along the
kink, i.e. perpendicular to the kink domain wall,
the energy of the latter per transverse volume $L^{d-1}$ is obtained
from summing/integrating zero-point energies according to
\bea\label{M1kdw}
{M^{(1)}\over L^{d-1}}&=&{m^3\over 3\lambda} + {1\over 2} \sum_B \int_{-\infty}^\infty 
{d^{d-1}\ell\over (2\pi)^{d-1}}
\sqrt{\omega_B^2+\ell^2}\nn\\&& + {1\over 2} \int_{-\infty}^\infty 
 {dk\,d^{d-1}\ell \over  (2\pi)^d }
\sqrt{k^2+\ell^2+m^2}\,\delta_K'(k)+\delta M
\eea
where the discrete sum is over the normalizable states $B$ of the
1+1-dimensional kink with
energy $\omega_B$, 
and the integral is over the continuum part of the
spectrum.

The spectrum of fluctuations for the 1+1-dimensional kink
is known exactly \cite{Raj:Sol}. It consists of a zero-mode,
a bound state with energy $\omega_B^2/m^2=3/4$, and scattering
states in a reflectionless potential for which
the phase shift $\delta_K(k)=-2\arctan(3mk/(m^2-2k^2))$ in the kink background
provides the difference in the spectral density, $\delta'_K(k)$,
between kink and trivial vacuum.

In a ``minimal'' renormalization scheme where tadpoles cancel but
$Z_\lambda=1$, one has
\be\label{dv2}
\delta M={3m\over 2\pi} 
d {\Gamma({-d\over 2})\over \Gamma(-\halb)(4\pi)^{d-1\over 2}}
\int_0^\infty dk (k^2+m^2)^{d-2\over 2},
\ee
yielding (with $x\equiv k/m$)
\bea\label{M1kdw2}
{M^{(1)}\over L^{d-1}}&=&{m^3\over 3\lambda} + {\Gamma({-d\over 2})\, m^d \over 
\Gamma(-\halb)(4\pi)^{d-1\over 2}} \biggl\{ \halb \left(3\over 4\right)^{d\over 2} \nn\\
&&\qquad\quad
+{3\over 4\pi} \int_{-\infty}^\infty dx (x^2+1)^{d-2\over 2}
\left[ {-1\over 4x^2+1}+(d-1) \right] \biggr\}.
\eea
Here the first term within the braces is the contribution from
the bound state with nonzero energy, and the second is the result
of combining the last two terms in (\ref{M1kdw}).

In the limit $d\to1$, which corresponds to the 1+1 dimensional kink,
one obtains
\be\label{M1DHN}
M^{(1)}_{d=1}=
{m^3\over 3\lambda} + \left( {m\over 4\sqrt3} - {3m\over 2\pi} \right),
\ee
reproducing the well-known DHN result \cite{Dashen:1974cj}.
It is interesting to note that it is the last term in (\ref{M1DHN})
that would be missed in a sharp-cutoff calculation (see Ref.
\cite{Rebhan:1997iv}) and that it
now arises from the last term in the square brackets of (\ref{M1kdw2}).

Eq.~(\ref{M1kdw2}) is also valid for $d\to2$ where it gives the
surface tension of a 2+1 dimensional kink domain wall; for higher
dimensions one has to include also a renormalization of the
coupling $\lambda$. All these results are in agreement with
those obtained previously by other methods \cite{Rebhan:2002uk}.

\subsection{One-loop susy kink mass}

Dimensional regularization is more delicate in susy theories.
To preserve susy, one should normally consider Siegel's dimensional
regularization by dimensional reduction \cite{Siegel:1979wq,Capper:1980ns}.
However, it is also possible to preserve susy by embedding
the susy kink in dimensions $\le 2+1$.

Embedding the susy kink in 2+1 dimensions gives a 
domain wall centered about a one-dimensional string on which
the fermion mass vanishes (since $U'(\ph_K)\propto \ph_K$ vanishes
at the center of the kink).
The total energy $M$ of the domain wall is infinite but the energy
density $M/L$ is finite; as a result there is strictly speaking no
zero mode in 2+1 dimensions associated with translational invariance. Indeed,
the zero mode of the kink is only normalizable in 1+1 dimensions, but
one can construct eigenfunctions in 2+1 dimensions which are products
of zero modes in 1+1 dimensions and plane waves in the orthogonal
direction(s) (along the domain wall). 
 
The 2+1 dimensional case is different also with respect to the
discrete symmetries of (\ref{Lss}). In 2+1 dimensions, 
$\gamma^5=\gamma^0\gamma^1\gamma^2 = \pm {\bf 1}$ corresponding
to the two inequivalent choices available for $\gamma^2=\pm\tau^1$
(in odd space-time dimensions the Clifford algebra has
two inequivalent irreducible representations).
Therefore, the sign of the fermion mass (Yukawa) term can no longer
be reversed by $\psi\to\gamma^5\psi$ and there is no longer
the $Z_2$ symmetry  $\ph\to-\ph, \psi\to\gamma^5\psi$.

What the 2+1 dimensional model does break spontaneously is instead
{\em parity}, which corresponds
to changing the sign of one of the spatial coordinates.
The Lagrangian is invariant under $x^m \to -x^m$ for
a given spatial index $m=1,2$ together with $\ph\to-\ph$ (which
thus is a pseudoscalar) and $\psi\to\gamma^m \psi$.
Each of the trivial vacua breaks these invariances spontaneously,
whereas a kink background in the $x^1$-direction with
$\ph_K(-x^1)=-\ph_K(x^1)$ is symmetric with respect to
$x^1$-reflections, but breaks $x^2=y$ reflection invariance.
 
This is to be contrasted with the 1+1 dimensional case, where
parity ($x^1\to-x^1$) can be represented either by $\psi\to\gamma^0\psi$
and a true scalar $\ph\to\ph$ or by $\psi\to\gamma^1\psi$ and
a pseudoscalar $\ph\to-\ph$. The former leaves the trivial vacuum
invariant, and the latter the ground state of the kink sector.

In what follows we shall consider the quantum corrections to
both, the mass of the susy kink and the tension of the
domain string, together. 
We again use a minimal renormalization scheme, where
inclusion of the fermionic tadpole loop simply replaces the
prefactor $3$ in (\ref{dv2}) by $(3-2)$.

In a Majorana representation of the Dirac matrices 
in terms of the usual Pauli matrices $\tau^k$
with $\g^0=-i\t^2$,
$\g^1=\t^3$, $\gamma^2=\tau^1$ (added for $d=2$), 
and $C=\t^2$ so that $\psi={\psi^+\choose\psi^-}$ with
real $\psi^+(x,t)$ and $\psi^-(x,t)$, the equations for the
bosonic and fermionic normal modes with frequency $\omega$ 
and longitudinal momentum $\ell$ (nonzero only when $d=2$) 
in the kink background $\ph=\ph_K$ read
\bea
&&[-\partial_x^2+U'{}^2+UU'']\eta=(\omega^2-\ell^2)\eta, \\
&&(\partial_x+U')\psi^++i(\omega+\ell)\psi^- = 0, \label{psip}\\
&&(\partial_x-U')\psi^-+i(\omega-\ell)\psi^+ = 0\label{psim}.
\eea
Acting with $(\partial_x-U')$ on (\ref{psip}) and eliminating
$\psi^-$ as well as $\ph'=-U$ shows that $\psi^+$ satisfies
the same equation as the bosonic fluctuation $\eta$.
Compared to $\psi^+$, the component $\psi^-$ has a continuous
spectrum whose modes differ by an additional phase shift
$\theta=-2\arctan(m/k)$ when traversing
the kink from $x_1=-\infty$ to $x_1=+\infty$, which is
determined only by $U'(\ph_K(x_1=\pm\infty))=
\pm m$.
Correspondingly, the difference of the spectral densities of the
$\psi^+$-fluctuations in the kink and in the trivial vacuum
equals that of the $\eta$-fluctuations, 
whereas that of $\psi^-$-fluctuations
is obtained by replacing $\delta'_K\to \delta'_K+\theta'$.

In the sum over zero-point energies
for the one-loop quantum mass of the kink (when $d=1$),
\be\label{Msumbf}
\tilde M=\tilde M_{cl.}+\halb\left(\sum\omega_B-\sum\omega_B'\right)
-\halb\left(\sum\omega_F-\sum\omega_F'\right)
+\d \tilde M \;,
\ee
one thus finds that the bosonic contributions from
the continuous spectrum are canceled by the fermionic
contributions\footnote{This cancellation could be however incomplete
for certain boundary conditions in global mode regularization.} 
except for the additional contribution
involving $\theta'(k)$ in the spectral density of the $\psi^-$ modes. 

The discrete bound states cancel
exactly, apart from the subtlety that the fermionic zero
mode should be counted as half a fermionic mode \cite{Goldhaber:2000ab}.
In strictly 1+1 dimensions, the zero modes do not contribute
simply because they carry zero energy, and for $d>1$, where
they become massless modes, they
do not contribute in dimensional regularization.

In a cutoff regularization in $d=2$, as we 
shall further discuss below,
they in fact do play a role. Remarkably, the half-counting of
the fermionic zero mode for $d=1$ has an analog for $d=2$ where the
bosonic and fermionic zero modes of the kink 
correspond to massless modes with energy
$|\omega|=|\ell|$. From (\ref{psip}) and (\ref{psim}) one
finds that the fermionic kink zero mode $\psi^+\propto \ph_K'$,
$\psi^-=0$
is a solution only for $\omega=+\ell$. It therefore cancels
only half of the contributions from the bosonic kink zero mode
which for $d=2$ have $\omega=\pm\ell$.
For $d=2$ one thus finds that the fermionic zero mode of the
kink corresponds to a chiral (Majorana-Weyl) fermion on
the domain wall (string) in 2+1 dimensions.
\cite{Callan:1985sa,Gibbons:2000hg}.%
\footnote{Choosing a different sign for $\gamma_1$
reverses the allowed sign of $\ell$ for these fermionic modes
and thus their chirality (with respect to the domain string
world sheet). This
corresponds to the other, inequivalent representation of the
Clifford algebra in 2+1 dimensions.}

In dimensional regularization, however, the kink zero modes and
their massless counterparts for $d>1$ can
be dropped, and the energy density of the
susy domain wall reads
\be\label{M1skdw}
{\tilde M^{(1)}\over L^{d-1}}={m^3\over 3\lambda} - \frac{1}{4} \int {dk\,d^{d-1}\ell \over  (2\pi)^d }
\sqrt{k^2+\ell^2+m^2}\,\theta'(k)+\delta \tilde M,
\ee
where
\be
\theta'(k)={2m\over k^2+m^2}.
\ee

With 
$\delta \tilde M = {1\over 3} \delta M $ the logarithmic
divergence in the integral in (\ref{M1skdw}) as $d\to1$
gets cancelled. A naive cut-off regularization at $d=1$ would
actually lead to a total cancellation of the $k$-integral with
the counter term $\delta \tilde M$, giving a vanishing
quantum correction in renormalization schemes with $\lambda=\lambda_0$.
In dimensional regularization there is now however a mismatch
for $d\not=1$ and a finite remainder in the limit $d\to1$
proportional to $(d-1)\Gamma(-(d-1)/2)$. The final result reads
\cite{Rebhan:2002uk}
\be\label{M1skdw2}
{\tilde M^{(1)}\over L^{d-1}}={m^3\over 3\lambda} 
-{m^d \over (4\pi)^{d+1\over 2}}{2\over d}\Gamma({3-d\over 2})\,.
\ee

In view of the discussion of the central charge below, it is
instructive to write the above finite remainder that dimensional
regularization leaves behind for $d\to1$ in the form
\be\label{eq:h22}
{\tilde M^{(1)}\over L^{d-1}}-{m^3\over 3\lambda}
=-{1\over 4}\int {dk\,d^{d-1}\ell\over (2\pi)^d}{\ell^2\over \omega}\theta'(k)
\ee
which is obtained by combining the integral in (\ref{M1skdw})
with the integral representation of the counter term (1/3 of the
r.h.s.~of (\ref{dv2})).
Evidently, the nonvanishing result is entirely due to the momenta
in the extra $d-1$ dimensions of a kink domain wall.

In the literature, at least to our knowledge,
only the case of a supersymmetric kink ($d=1$) 
has been considered and dimensional regularization reproduces
the result obtained before by 
Refs.~\cite{Schonfeld:1979hg,Boya:1988zh,Nastase:1998sy,Graham:1998qq,Shifman:1998zy,Bordag:2002dg}.

However, a (larger) number of papers have missed
the contribution $-m/(2\pi)$, mostly because of the (implicit) use
of an inconsistent energy-cutoff scheme
\cite{Kaul:1983yt,Imbimbo:1984nq,Chatterjee,Yamagishi:1984zv}
or have obtained different answers because of the use
of boundary conditions that accumulate a finite amount of
energy at the boundaries \cite{Uchiyama:1986gf,Rebhan:1997iv}.
The former result is however now generally accepted and,
in the case of the super-sine-Gordon model (where the
same issues arise with the same results) in agreement with
S-matrix factorization \cite{Ahn:1991uq}.

A new result, which follows from (\ref{M1skdw2}) and which
will play a role for the discussion of central
charges in the next section, is the nonvanishing one-loop correction 
\be
{\tilde M^{(1)}_{d=2}\over L}=-{m^2\over 8\pi}
\ee
for the surface tension of the minimally
susy kink domain wall in 2+1 dimensions.

In Ref.~\cite{Litvintsev:2000is} the correct susy kink mass
has also been obtained by employing a smooth energy (momentum) cutoff,
the necessity of which becomes apparent, as in the purely bosonic
case, by considering the 2+1 dimensional domain wall.
Using a naive cutoff for $d=2$ one finds quadratic divergences
which cancel only upon inclusion of the zero modes (which become massless
modes in 2+1 dimensions). As we have
discussed above, unlike the other bound states, these do not
cancel because the fermionic zero mode becomes a chiral
fermion on the domain-string world-sheet and thus 
cancels only half of the bosonic zero (massless) mode contribution,
yielding
\bea
&&\int_0^\infty {d\ell\over 2\pi} \biggl\{\halb\sqrt{\ell^2}
-\int_{-\Lambda_k}^{\Lambda_k}{dk\over 2\pi} 
\left[\sqrt{k^2+\ell^2+m^2}{m\over k^2+m^2}-{1\over \sqrt{k^2+\ell^2+m^2}} \right]
\biggr\}\nn\\
&&
\stackrel{\Lambda_k\to\infty}{\longrightarrow}
\int_0^\infty{d\ell\over 2\pi} \biggl\{{\ell\over 2}-{\ell\over \pi}
\arctan{\ell\over m}\biggr\}
\sim\int_0^\infty{d\ell\over \pi}{m\over 2\pi}
\eea
which is however still linearly divergent. Smoothing out the
cutoff in the $k$-integral does pick an additional (and for
$d=1$ the only) contribution $-m/(2\pi)$, which is now
necessary to have a finite result for $d=2$. This finite
result then reads
\be
{\tilde M^{(1)}_{d=2}\over L}=-{1\over \pi}\int_0^\infty {d\ell\over 2\pi}
\left(m-\ell\arctan{m\over \ell}\right)=-{m^2\over 8\pi}
\ee
in agreement with the result obtained above in dimensional regularization.

\subsection{Susy algebra and its quantum corrections}

\subsubsection{Dimensional regularization}

The susy algebra for the 1+1 and the 2+1 dimensional cases can
both be covered by starting from 2+1 dimensions, the 1+1 dimensional
case following from reduction by one spatial dimension.


In 2+1 dimensions one
obtains classically \cite{Gibbons:1999np}
\begin{eqnarray}
  \label{eq:3dsusy}
 \{Q^{\alpha},\bar{Q}_{\beta}\}&=&2i(\gamma^M)^{\alpha}{}_{\beta}P_M\ ,
          \quad (M=0,1,2)\nonumber\\
    &=&2i(\gamma^0H+\gamma^1(\tilde{P}_x+\tilde{Z_y})
    +\gamma^2(\tilde{P}_y-\tilde{Z}_x))^\alpha{}_\beta,
\end{eqnarray}%
where we separated off two surface terms $\tilde Z_m$ in defining
\begin{eqnarray}
  \label{eq:Ptilde}
\tilde P_m = \int d^dx  \tilde{\mathcal{P}}_m, \quad 
  &&\tilde{\mathcal{P}}_m=\dot\ph\,\partial_m\ph
      -\halb(\bar{\psi}\gamma^0\partial_m\psi),\\
  \label{eq:Ztilde}
\tilde Z_m = \int  d^dx  \tilde{\mathcal{Z}}_m, \quad
  &&\tilde{\mathcal{Z}}_m=U(\ph) \partial_m\ph = \partial_m W(\ph)
\end{eqnarray}
with $W(\ph)\equiv\int d\ph\, U(\ph)$.

Having a kink profile in the $x$-direction, which satisfies the
Bogomolnyi equation $\partial_x \ph_K=-U(\ph_K)$, one finds
that with our choice of Dirac matrices
\bea
  \label{eq:qpm}
  &&Q^{\pm}=\int d^2x[(\dot\ph\mp\partial_y\ph)\psi^\pm
      +(\partial_x\ph\pm U(\ph))\psi^\mp],\\
&&\{Q^\pm,Q^\pm\}=2(H \pm (\tilde Z_x - \tilde P_y)),
\eea
and the charge $Q^+$ 
leaves the topological (domain-wall) vacuum $\ph=\ph_K$, $\psi=0$
invariant.
This corresponds to classical
BPS saturation, since with $P_x=0$ and $\tilde P_y=0$
one has $\{Q^+,Q^+\}=2(H+\tilde Z_x)$ and, indeed, with a kink domain wall
$\tilde Z_x/L^{d-1}=W(+v)-W(-v)=-M/L^{d-1}$.

At the quantum level, hermiticity of $Q^\pm$ implies
\be\label{HPyineq}
\langle s|H|s\rangle  \ge |\langle s|P_y|s\rangle |  \equiv |\langle s|(\tilde P_y-\tilde Z_x)|s\rangle |.
\ee
This inequality is saturated when
\be
Q^+|s\rangle =0
\ee
so that BPS states correspond to massless states $P_M P^M=0$
with $P_y=M$ for a kink domain wall in the $x$-direction \cite{Losev:2000mm},
however with infinite momentum and energy unless the $y$-direction
is compact with finite length $L$.
An antikink domain wall has instead $ Q^-|s\rangle =0$. In both cases,
half of the supersymmetry is spontaneously broken.


Classically, the susy algebra in 1+1 dimensions is obtained from
(\ref{eq:3dsusy}) simply by dropping $\tilde P_y$
as well as $\tilde Z_y$ so that $P_x\equiv\tilde P_x$.
The term $\gamma^2 \tilde Z_x$ remains, however, with
$\gamma^2$ being the nontrivial $\gamma^5$ of 1+1 dimensions.
The susy algebra simplifies to
\be
\{Q^\pm,Q^\pm\} = 2(H\pm Z),\quad \{Q^+,Q^-\}=2P_x
\ee
and one has the inequality
\be
\langle s|H|s\rangle  \ge |\langle s|Z|s\rangle |
\ee
for any quantum state $s$. BPS saturated states have
$Q^+|s\rangle =0$ or $Q^-|s\rangle =0$, corresponding to
kink and antikink, respectively, and break half of the
supersymmetry.

In a kink (domain wall) background with only nontrivial $x$ dependence,
the central charge density $\tilde \mathcal Z_x$ receives
nontrivial contributions.
Expanding $\tilde \mathcal Z_x$ 
around the kink background gives
\begin{eqnarray}
  \label{eq:Znaiv}
  \tilde{\mathcal{Z}}_x=U\partial_x\ph_K-\frac{\delta\mu^2}{\sqrt{2\lambda}}
        \partial_x\ph_K+\partial_x(U\eta)+\halb\partial_x(U{'}\eta^2)
        +O(\eta^3). 
\end{eqnarray}
where only the part quadratic in the fluctuations contributes to
the integrated quantity at one-loop order\footnote{But this does not hold
for the central charge density locally 
\cite{Shifman:1998zy,Goldhaber:2001rp}.}.
However, this matches
precisely the counter term $\delta M$ from requiring vanishing tadpoles.
Straightforward application of the rules of
dimensional regularization thus leads to a null result for
the net one-loop correction to $\langle\tilde Z_x\rangle $ in the same way
as found in Refs.~\cite{Imbimbo:1984nq,Chatterjee,Rebhan:1997iv,Nastase:1998sy}
in other schemes.

On the other hand, by considering the less singular
combination $\langle H+\tilde Z_x\rangle $ and showing that it vanishes exactly, 
it was concluded in Ref.~\cite{Graham:1998qq}
that $\langle\tilde Z_x\rangle $ has to compensate any nontrivial result
for $\langle H\rangle $, which in Ref.~\cite{Graham:1998qq} was obtained
by subtracting successive Born approximations for scattering
phase shifts. In fact, Ref.~\cite{Graham:1998qq} explicitly
demonstrates how to rewrite $\langle\tilde Z_x\rangle $ into $-\langle H\rangle $,
apparently without the need for the anomalous terms in the quantum central
charge operator
postulated in Ref.~\cite{Shifman:1998zy}.

The resolution of this discrepancy is that Ref.~\cite{Graham:1998qq}
did not regularize $\langle\tilde Z_x\rangle $ and 
therefore the manipulations
needed to rewrite it as $-\langle H\rangle $ (which eventually is
regularized and renormalized) are ill-defined.
Using dimensional regularization naively one in fact obtains
a nonzero result for $\langle H+\tilde Z_x\rangle $, apparently
in violation of susy.

However, dimensional regularization by embedding the kink
as a domain wall in (up to) one higher dimension, which
preserves susy, instead leads to 
\be
\langle H+\tilde Z_x-\tilde P_y\rangle =0,
\ee
i.e. the saturation of (\ref{HPyineq}), as we shall now verify.

The bosonic contribution to $\langle\tilde P_y\rangle $ involves
\begin{equation}
  \label{eq:pybos}
  \halb\langle\dot{\eta}\partial_y\eta+\partial_y\eta\dot{\eta}\rangle=
  -\int\frac{d^{d-1}\ell}{(2\pi)^{d-1}}\intsum {dk\over 2\pi}\ 
\frac{\ell}{2}|\phi_k(x)|^2,
\end{equation}
where the $\phi_k(x)$ are the mode functions of the
fluctuation field operator $\eta$.
The $\ell$-integral factorizes and gives zero both because it is
a scale-less integral and because the integrand is odd in $\ell$.

The fermions on the other hand turn out to give nontrivial contributions:
The mode expansion for the fermionic field operator reads
\begin{eqnarray}
  \label{eq:ferm}
 \psi\!&=&\!\psi_0+\int\frac{d^{d-1}\ell}{(2\pi)^{\frac{d-1}{2}}}
\intsum'\frac{dk}{\sqrt{4\pi\omega}}
 \left[b_{k,\ell}\  e^{-i(\omega t-\ell y)}
      { {\scriptstyle\sqrt{\omega+\ell}}\ \phi_k(x)\choose
                    {\scriptstyle\sqrt{\omega-\ell}}\ is_k(x)}
            + b^{\dagger}_{k,\ell}\ (c.c.)\right],\nonumber\\
 &&\psi_0=\int\frac{d^{d-1}\ell}{(2\pi)^{\frac{d-1}{2}}}b_{0,\ell}\ e^{-i\ell (t- y)}
   {\phi_0\choose 0},\quad b^{\dagger}_0(\ell)=b_0(-\ell),
\end{eqnarray}
where $\psi_0$ is the fermionic zero-mode lifted to a Majorana-Weyl
domain wall fermion, and
$s_k=\frac{1}{\sqrt{\omega^2-\ell^2}}(\partial_x+U{'})\phi_k$.
This leads to
\begin{eqnarray}
  \label{eq:py}
  \langle\tilde{\mathcal{P}}_y\rangle&=&
  \frac{i}{2}\langle\psi^{\dagger}\partial_y\psi\rangle\nonumber\\
&=& \halb \int\frac{d^{d-1}\ell}{(2\pi)^{d-1}}\intsum {dk\over 2\pi}{\ell\over 2\omega}
 \left[(\omega+\ell)|\phi_k|^2+(\omega-\ell)|s_k|^2\right]\nonumber\\
  &=&\halb\int\frac{d^{d-1}\ell}{(2\pi)^{d-1}}\ \ell\ \theta(-\ell)\ |\phi_0|^2+
              \nonumber\\
  &&+\halb\int\frac{d^{d-1}\ell}{(2\pi)^{d-1}}\intsum' {dk\over 2\pi}
   \left( \frac{\ell}{2}(|\phi_k|^2+|s_k|^2)
    +\frac{\ell^2}{2\omega}(|\phi_k|^2-|s_k|^2)
\right).\qquad
\end{eqnarray}
{}From the last sum-integral we have separated off the contribution
of the zero mode of the kink (the chiral domain wall fermion
for $d>1$). The contribution of the latter no longer vanishes by
symmetry, but the $\ell$-integral is still scale-less and therefore
put to zero in dimensional regularization. The first sum-integral
on the right-hand side is again zero by both symmetry and scalelessness,
but the final term is not: The $\ell$-integration no longer
factorizes because $\omega=\sqrt{k^2+\ell^2+m^2}$, and
leads to a nonvanishing result, which, as one can show \cite{Rebhan:2002yw},
is identical to 
the finite net contribution
in $\langle\mathcal H\rangle $.
For the integrated quantities, this equality
can be seen by comparing with (\ref{eq:h22})
upon using that $\int dx(|\phi_k|^2-|s_k|^2)=\theta'(k)$.

So for all $d\le2$ we have BPS saturation, $\langle H\rangle =|\langle\tilde Z_x-\tilde P_y\rangle |$,
which in the limit $d\to1$, the susy kink, is made possible by 
a nonvanishing $\langle\tilde P_y\rangle $. The anomaly in the central charge
is seen to arise from a parity-violating contribution in $d=1+\epsilon$
dimensions which is the price to be paid for preserving (minimal) supersymmetry
when going up in dimensions to embed the susy kink as a domain wall.



To summarize, in 2+1 dimensions, we have $P_y=\tilde P_y-\tilde Z_x$
and $|\langle P_y\rangle |=\langle H\rangle $, where $\tilde P$ and $\tilde Z$ were
defined in (\ref{eq:3dsusy}). Classically, this BPS saturation
is guaranteed by $\tilde Z_x$ alone. At the quantum level,
however, the quantum corrections to the latter are cancelled
completely by the counter term from renormalizing tadpoles to zero.
All nontrivial corrections come from the ``genuine'' momentum
operator $\tilde P_y$, and are due to having a spontaneous
breaking of parity.

In the limit of 1+1 dimensions, because $\gamma^2|_{D=2+1}=
\gamma^5|_{D=1+1}$, one has to make the
identification $Z=\tilde Z_x-\tilde P_y$. For $\tilde Z_x$, one
again does not obtain net quantum corrections. However, the
expectation value $\langle\tilde P_y\rangle $ does not vanish in the
limit $d\to1$, although there is no longer an extra dimension.
The spontaneous parity violation in the 2+1 dimensional theory, which had to be
considered in order to preserve susy, leaves a finite imprint upon
dimensional reduction to 1+1 dimensions by providing
an anomalous additional contribution to $\langle\tilde Z_x\rangle $ balancing
the nontrivial quantum correction $\langle H\rangle $. 

\subsubsection{Dimensional reduction}

We now show how the central charge anomaly can be recovered
from Siegel's version of dimensional regularization 
\cite{Siegel:1979wq,Capper:1980ns}
where $n$ is smaller than the dimension of spacetime and where one keeps
the number of field components fixed, but lowers the number of
coordinates and momenta from 2 to $n<2$. At the one-loop level one
encounters 2-dimensional $\delta_\mu^\nu$ coming from
Dirac matrices, and $n$-dimensional $\hat\delta_\mu^\nu$ from
loop momenta. An important concept which is going to play a role is that of
the evanescent counter\-terms \cite{Bonneau}
involving the factor ${1\over \epsilon}\hat{\hat{\delta}}{}_\mu^\nu$, 
where $\hat{\hat\delta}{}_\mu^\nu\equiv \delta_\mu^\nu-
\hat\delta_\mu^\nu$ has only $\epsilon=2-n$ nonvanishing components.

Consider now the supercurrent 
$j_\mu=-(\not\!\partial\ph+U(\ph))\gamma_\mu\psi$.
In the trivial vacuum, expanding into quantum fields yields
\be\label{jmuexp}
j_\mu=-\left(\not\!\partial\eta+U'(v)\,\eta
+
\halb U''(v)\,\eta^2
\right)\gamma_\mu
\psi + {1\over \sqrt{2\lambda}}\delta\mu^2\gamma_\mu\psi,
\ee
where $v=\mu/\sqrt\lambda$.
Only matrix elements with one external fermion are divergent.
The term involving $U''(v)\eta^2$ in (\ref{jmuexp}) gives rise to
a divergent scalar tadpole that is cancelled 
completely by the counter term $\delta\mu^2$ (which
itself is due to an $\eta$ and a $\psi$ loop). The only other
divergent diagram is due to the term involving $\not\!\partial\eta$
in (\ref{jmuexp}) and has the form of a $\psi$-selfenergy. Its
singular part reads
\be
\langle0| j_\mu |p\rangle ^{\rm div} = i U''(v)
\int_0^1 dx \int {d^n \kappa\over (2\pi)^n} {\not\!\kappa \gamma_\mu\! \not\!\kappa
\over [\kappa^2 + p^2 x(1-x) + m^2]^2}u(p).
\ee
Using $\hat\delta_\mu^\nu\equiv \delta_\mu^\nu-\hat{\hat\delta}{}_\mu^\nu$
we find that under the integral
$$\not\!\kappa \gamma_\mu\!\! \not\!\kappa = - \kappa^2(\delta_\mu^\lambda-
{2\over n}\hat \delta_\mu^\lambda)\gamma_\lambda=
{\epsilon\over n}\kappa^2\gamma_\mu - {2\over n}\kappa^2
\hat{\hat\delta}{}_\mu^\lambda \gamma_\lambda$$
so that
\be
\langle0| j_\mu |p\rangle ^{\rm div} = {U''(v)\over 2\pi}{\hat{\hat\delta}{}_\mu^\lambda
\over \epsilon} \gamma_\lambda u(p).
\ee
Hence, the regularized one-loop contribution to the susy 
current contains the evanescent operator
\be\label{jsusydiv}
j_\mu ^{\rm div} = {U''(\ph)\over 2\pi}{\hat{\hat\delta}{}_\mu^\lambda
\over \epsilon} \gamma_\lambda  \psi.
\ee
This is by itself a conserved quantity, because all fields
depend only on the $n$-dimensional coordinates.
The renormalized susy current
$j_\mu^{\rm ren.}=j_\mu-j_\mu^{div}$
is thus still conserved,%
\footnote{Note also that (\ref{jsusydiv}) does not change the
susy charges $Q=\int dx j_0$ if one assumes that $\hat{\hat\delta}{}_\mu^\nu$
has only spatial components.
Furthermore, recall that conserved currents do not renormalize.}
but from the evanescent counter term it receives a nonvanishing
contribution to $\gamma\cdot j^{\rm ren.}$ which
appears in the divergence of the renormalized 
conformal-susy current $j_\mu^{\rm ren.}$
\be
\partial^\mu (\not\!x j_\mu^{\rm ren.})_{\rm anom.}=-
\gamma^\mu j_\mu^{\rm div}= -{U''\over 2\pi}\psi.
\ee
(There are also nonvanishing
nonanomalous contributions
to $\partial^\mu (\mbox{$\not\!x$} j_\mu)$ because our model is not
conformal-susy invariant at the classical level.)

Ordinary susy on the other hand is unbroken; there is no anomaly
in the divergence of $j_\mu^{\rm ren.}$. A susy variation of
$j_\mu$ involves the energy-momentum tensor and the
topological central-charge current $\zeta_\mu$
according to
\be
\delta j_\mu = -2T_\mu{}^\nu \gamma_\nu \epsilon - 2 \zeta_\mu \gamma^5 
\epsilon,
\ee
where classically $\zeta_\mu=\epsilon_{\mu\nu}U
\partial^\nu \varphi$.

At the quantum level,
the counter-term $j_\mu^{\rm ct}=-j_\mu^{\rm div}$ induces
an additional contribution to the central charge current
\be
\zeta_\mu^{\rm anom}={1\over 4\pi}
{\hat{\hat\delta}{}_\mu^\nu\over \epsilon}
\epsilon_{\nu\rho}\partial^\rho U'
\ee
which despite appearances is a {\em finite} quantity: using that total
antisymmetrization of the three lower indices has to vanish
in two dimensions gives
\be
\hat{\hat\delta}{}_\mu^\nu
\epsilon_{\nu\rho} = \epsilon \epsilon_{\mu\rho}+
\hat{\hat\delta}{}_\rho^\nu
\epsilon_{\nu\mu} 
\ee
and together with the fact the $U'$ only depends on $n$-dimensional
coordinates this finally yields
\be\label{zetaanom}
\zeta_\mu^{\rm anom}={1\over 4\pi} \epsilon_{\mu\rho} \partial^\rho U'
\ee
in agreement with the anomaly in the central charge as obtained
previously.\footnote{It would be interesting to study further
the infrared/ultraviolet connection for this anomaly.}

We emphasize that $\zeta_\mu$ itself does not require the
subtraction of an evanescent counterterm. The latter only
appears in the susy current $j_\mu$, which gives rise
to a conformal-susy anomaly in $\not \!x j_\mu$.
A susy variation of the latter
shows that it forms a conformal current multiplet involving besides
the dilatation current $T_{\mu\nu}x^\nu$ and the Lorentz current
$T_\mu{}^\nu x^\rho\epsilon_{\nu\rho}$
also a current 
\be\label{cccc}
j_{(\nu)}^{(\zeta)\mu}=x^\rho \epsilon_{\rho\nu}\zeta^\mu.
\ee
We identify this with the conformal central-charge current, which
is to be distinguished from the ordinary central-charge current $\zeta_\mu$.

The anomalous contribution to the ordinary central charge is thus
understood as the additional nonconservation at the quantum level
of the {\em conformal} central-charge current (\ref{cccc}).
(Additional, because the model is not conformally invariant so that
there is already nonconservation at the classical level.)
This finally answers the question: what kind of anomaly corresponds
to the anomalous contribution to the central charge?

\section{Local Casimir energy for solitons}

We have seen in the introduction that 
there is a problem with the regularization of the zero-point
energies by means of mode number cutoff (equal numbers of modes in each sector, 
with careful counting of zero modes): it includes spurious boundary energy.
On the other hand, the principle of mode regularization seems natural,
so the question arises whether we can devise a mode number cutoff
scheme without the unwanted boundary energy. This almost automatically leads
to a new regularization scheme for Casimir sums, 
called local mode regularization.
Given that each mode determines a mode function $\phi_n(x,t)$ (or $\psi_n(x,t)$
for the fermions) normalized such that
for a large box with volume $L$ the $\phi_n(x,t)$ become
at large $|x|$ a 
plane wave with unit strength (the corrections to these plane waves are of 
order $L^{-1/2}$),
we can introduce a concept of local mode density $\rho(x)$ in the kink 
sector (and $\rho^{(0)}(x)$ in the trivial sector) as follows:  
$\rho(x)=\sum\phi_n^\ast(x)\phi_n(x)$.
To regulate such sums we would like to again cut off the 
sum over $n$ at a large number $N$.

The kink mass contains the difference of the energy sum 
$\sum|\phi_n(x)|^2\hal\omega_n$ in the kink sector and the energy sum 
$\sum|\phi_n^{(0)}(x)|^2\hal\omega_n^{(0)}$ in the trivial sector. The problem 
is thus how to relate the regularization in one sector to that in the other
sector. The most straightforward method would be to include the same number of
terms in each sum, just as in the case of global mode number regularization.
However in this way the sums would only indirectly take the presence of the kink 
into account (through the inequality of the $\omega_n$ and $\omega_n^{(0)}$).

We now formulate a principle which we have not yet been able to prove that it
is equivalent to other principles, or that it preserves supersymmetry, but which
gives correct answers for the kink mass and supersymmetric kink mass, and which is so simple that it deserves 
further study. Namely we require that the regulated mode densities in both sectors
are equal. The function $\sum_{n=0}^N\phi_n^\ast(x)\phi_n(x)$
is a function of $N$ or equivalently of $\Lambda=2\pi N/L$, but for large $L$ 
we can interpolate it to become a function of a continuous variable $\Lambda$. 
Similarly, $\rho_N^{(0)}$ becomes a continuous function of $\Lambda$. Since 
$L^{-1}\int|\phi_k|^2dx=1$ counts each mode once, 
it may seem natural to also use 
$|\phi_k(x)|^2$ to count modes locally. However note that 
$\left<\vf(x)\vf(x)\right>$ contains $\sum|\phi_k(x)|^2\frac{1}{2\omega_k}$ while 
the energy density contains $\sum|\phi_k(x)|^2\hal{\omega_k}$. The choice to use
$\sum|\phi_k(x)|^2$ to define a regularization is perhaps natural, certainly 
more natural than for example $\sum|\phi_k(x)|^4$, but we have not proven that 
$\sum|\phi_k(x)|^2$ is the correct object. 

If the density $\rho(x)$ is cut off
at a large $\Lambda$, the density $\rho^{(0)}(x)$ should be cut off at a 
$\Lambda+\Delta\Lambda\equiv 2\pi N^{(0)}$ such that $\rho_{\Lambda}(x)$ is equal
to $\rho^{(0)}_{\Lambda+\Delta\Lambda}(x)$. Far away from the kink all modes are 
plane waves, so for large $|x|$ one expects $\Delta\Lambda$ to vanish, but near 
the kink $\Delta\Lambda$ will be nonvanishing. This implies that 
$\Delta\Lambda$ is $x$-dependent, and the principle of local mode regularization 
takes the following form
\begin{equation}
  \label{eq:lmp}
  \rho_{\Lambda}(x)=\rho^{(0)}_{\Lambda+\Delta\Lambda(x)}(x).
\end{equation}
The regulated energy densities in the kink and the trivial sector, given by
  $\ve=\sum|\phi_n(x)|^2\hal\omega_n$ and 
    $\ve^{(0)}=\sum|\phi_n^{(0)}(x)|^2\hal\omega_n^{(0)}$,
will then in general be different if the regulated densities $\rho(x)$ and 
$\rho^{(0)}(x)$ are equal.

It is now straightforward to calculate the local Casimir mass of a soliton. It is 
given by
{\setlength\arraycolsep{0pt}
\begin{eqnarray}
  \label{eq:lmass}
  &&\epsilon_{\rm Cas}(x)= \epsilon(x) - \epsilon^{(0)} (x) = \nonumber\\
   &&\frac{1}{2} \omega_B \phi_B^2(x) +  2\int_0^\Lambda \frac{dk}{2 \pi} 
    |\phi(k,x)|^2 \frac{1}{2} \omega - 2 \int_0^{\Lambda+\Delta \Lambda(x)}
     \frac{dk}{2\pi}  \frac{1}{2} \omega + \delta M(x).
\end{eqnarray}}
The bound state yields a zero-point energy $\hal\omega_B$ and has (normalizable) 
mode function $\phi_B(x)$, while the continuous spectrum in the kink sector 
consists of plane waves $\phi^{(0)}(k,x)=\exp{ikx}$. We rewrite this expression 
such that it is manifestly finite
{\setlength\arraycolsep{0pt}
\begin{eqnarray}
  \label{eq:ecasf}
  &&\ve_{\rm Cas}(x)= \hal\omega_B\phi_B^2(x)\nonumber\\
  &&+\left\{2\int_0^{\Lambda}\frac{dk}{2\pi}
    \left(|\phi(x,k)|^2-1\right)\hal\omega+\delta M(x)\right\}-
    \frac{\Delta\Lambda(x)}{2\pi}\Lambda.
\end{eqnarray}}
The last term is the "anomaly", it appears here as a term of the form 
$\Lambda/\Lambda$ because $\Delta\Lambda(x)$ is proportional to $1/\Lambda$
as well see presently.

In the kink 
sector $\phi(k,x)$ can be given explicitly. From the explicit form one finds 
that it can be expressed in terms of the wave functions of the discrete 
spectrum as follows
\begin{equation}
  \label{eq:dis}
   |\phi(k,x)|^2 - 1 = - \sum_j \phi_{j}^2(x)
    \frac{2\sqrt{m^2-\omega_{j}^2}}{\omega^2-\omega^2_{j}}.
\end{equation}
(For the kink $j$ refers to the zero mode with $\omega^2_j=0$ and the bound state 
with $\omega_j^2=3/4m^2$.)
This formula seems to be new and we interpret it as a 
local version of the completeness relation. Integration over $k$ yields the usual
completeness relation
\begin{equation}
  \label{eq:comp}
  \int_{-\infty}^{\infty} \frac{dk}{2 \pi} \left\{|\phi(k,x)|^2-1\right\}+ 
    \phi_0^2(x) + \phi_B^2(x) =0,
\end{equation}
but the local version allows us to evaluate $\Delta\Lambda$ as
\bea
\Delta\Lambda(x)&=&\int_\Lambda^\infty dk\, \left(1-|\phi(k,x)|^2\right)=
2\int_\Lambda^\infty\!\! dk\,\sum_j {\sqrt{m^2-\omega_j^2} \over \omega^2-\omega_j^2}
\phi_j^2(x)\nn\\&=&
{3m^2\over 4\Lambda \cosh{mx\over 2}}+\mathcal O(\Lambda^{-2}).
\eea

The local counter term $\delta M(x)$ is of course 
equal to the term proportional 
to $\delta\mu^2$ in the energy density.%
\footnote{It should also cancel the divergence in the integral
$\int_0^\infty \frac{dk}{2\pi}(|\phi(k,x)|^2-1)\hal\omega$. This yields another
amusing formula,
$$  \frac{\sum_j2\phi^2_j(x)
\sqrt{m^2-\omega_j^2}}{\sum_j2\sqrt{m^2-\omega^2_j}}=
    \frac{\vf_K^2(x)-\vf_K^2(\infty)}{\int dx\left(\vf_K^2(x)-\vf_K^2(\infty)\right)}.
$$
} We can now substitute all these relations and find then
\begin{equation}
  \label{eq:ecasres1}
  \epsilon_{\rm Cas}(x) = \left(\frac{1}{2} \omega_B - \frac{m}{2 \sqrt{3}} -
     \frac{m}{2 \pi}\right)\phi_B^2(x) - \frac{m}{\pi} \phi_0^2(x).
\end{equation}
We can rewrite this formula as
\begin{equation}
  \label{eq:ecasres2}
  \epsilon_{\rm Cas}(x)= \sum_j \frac{1}{2} \left( 1 - \frac{2}{\pi} \arctan 
    \frac{\omega_j}{\sqrt{m^2- \omega_j^2}} \right)\omega_j\phi^2_{j}(x)-
   \sum_j\frac{1}{\pi}\sqrt{m^2-\omega^2_{j}}\phi_{j}^2(x).
\end{equation}
Such expressions are known for the total energy, but this local version 
seems new.

The local Casimir energy is not, however, equal to the local energy density. 
There are two further terms:

(i) the energy density contains a term $\hal(\partial_x\eta)^2$ (where 
$\eta(x,t)$ is the fluctuation field), but Casimir energies contain eigenvalues 
of the field operator which contains a term $-\partial_x^2\vf$ and our local
Casimir energy gets contributions from $-\hal\eta\partial_x^2\eta$. The 
difference, denoted by $\Delta\ve_{\rm Cas}(x)$, is a double total derivative
\begin{equation}
  \label{eq:decas}
  \Delta\ve_{\rm Cas}(x)=
\left<\hal \partial_x \eta \partial_x \eta \right>
-\left<-\hal \eta \partial_x^2 \eta \right>=
\textstyle{\frac{1}{4}}\partial_x^2\left<\eta(x)\eta(x)\right>.
\end{equation}
The propagator $\left<\eta(x)\eta(y)\right>$ contains a singularity as $x$
 tends $y$, but this singularity is $x$-independent and cancels due to the 
space derivatives. So $\Delta\ve_{\rm Cas}(x)$ is a finite and smooth function;

\begin{figure}
\centerline{
\includegraphics[bb=100 600 235 675]{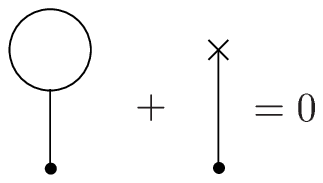}
\includegraphics[bb=100 600 235 675]{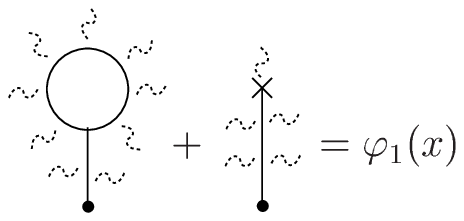}
}
\caption{Renormalized tadpoles in the vacuum and in the kink
background, respectively. 
\label{figphi1}}
\end{figure}

(ii) near the kink, the propagators of $\eta$ are deformed: they become 
(complicated) expressions for propagation in a kink background. Thus the cancellation
 of tadpoles which we imposed in flat space and which gave us the mass 
renormalization $\mu_0^2=\mu^2+\delta\mu^2$, no longer holds in the vicinity 
of the kink. Instead, one has in the kink sector
\begin{equation}
  \label{eq:fi1}
  \vf(x,t)=\vf_K(x,t)+\vf_1(x,t)+\eta(x,t),
\end{equation}
where $\left<\eta(x,t)\right>=0$ by definition, and $\vf_1(x)$ is a mean field
induced by the kink \cite{Shifman:1998zy}. 
This mean field gives another contribution to the energy 
density which we denote by $\Delta\ve_{(\vf_1)}(x)$ and which 
follows from expanding $\left<\mathcal E(x)\right>=\halb
\left< \partial_x \vf \partial_x \vf \right>+\halb \left< U^2\right>$,
\begin{equation}
  \label{eq:ecasf1}
  \Delta \epsilon_{(\varphi_1)}(x)=\partial_x\varphi_1
    \partial_x\varphi_K+(\frac{1}{2}U^2)'\varphi_1=
      \partial_x(\varphi_1\partial_x\varphi_K).
\end{equation}

The field $\vf_1(x)$ follows from the vanishing of the expectation value
of the field equation of the Heisenberg fields
$\Phi$. Using $\left<\eta(x,t)\right>=0$
one easily obtains
{\setlength\arraycolsep{2pt}
\begin{eqnarray}
  \label{eq:Heom}
  &&\langle -\partial_t^2\Phi +\partial_x^2\Phi-(\frac{1}{2}U^2)'\rangle=0 \\
  &&=\partial_x^2\varphi_1-(\frac{1}{2}U^2)''\varphi_1- \frac{1}{2!}(\frac{1}{2}U^2)'''
     \langle \eta^2\rangle-\frac{1}{2}\delta m^2\varphi_K
\end{eqnarray}}
The sum of the last two terms is again smooth and finite, and if we rewrite this 
equation as
\begin{eqnarray}
  \label{eq:f1res}
  \vf_1(x)=\left[\partial_x^2-(\hal U^2){''}\right]^{-1}\langle\eta^2(x)
            -\eta^2(\infty)\rangle 3\lambda\vf_K(x)
\end{eqnarray}
we recognise the Feynman graphs we depicted in Fig.~\ref{figphi1}.

The solution for $\vf_1(x)$ is of the form $Ax\eta_0(x)+ 
B\partial_x\eta_0(x)$ (note that $\left<\eta^2(x)\right>
-\left<\eta^2(\infty)\right>$ contains terms with $\cosh^{-4}(mx/2)$ and 
$\eta_0\sim\cosh^{-2}(mx/2)$, and the fluctuation operator 
$\partial_x^2-(\hal U^2){''}$ vanishes on $\eta_0$). The term 
$Ax\eta_0(x)$ can also be written as proportional to $\vf_K(x(1+A))$ because 
$\eta_0(x)\sim\partial_x\vf_K$, and this rescaling of $x$ can also be written 
as a rescaling of $m$ (since $\vf_K$ depends only on $\hal mx$) and a counter 
rescaling of $\lambda$ to keep the prefactor $\mu/\sqrt{\lambda}$ invariant:
\begin{eqnarray}
  \label{eq:a}
  Ax\eta_0(x)\sim A(m\frac{\partial}{\partial m}
     +2\lambda\frac{\partial}{\partial\lambda})\vf_K(x).
\end{eqnarray}
One can also write this as 
$\vf_K(m,\lambda,x)+Ax\eta_0(x)
=\vf_K(\bar{m},\bar{\lambda},x)$ where we discover that
this rescaling of the renormalized mass $m$ yields the pole mass $\bar{m}$
\cite{Shifman:1998zy}!
We have
not been able to give a similarly simple physical explanation of the rescaled 
coupling $\bar{\lambda}=(\bar{m}/m)^2\lambda$.

One can now substitute all expressions to get explicit formulas for 
the complete energy density ${\mathcal E}(x)$
for the kink (or for any other 1+1 dimensional soliton). 
One can also repeat this 
exercise for the supersymmetric kink (in this case the only difference 
for $\varphi_1$ is a different result for $\bar{m}$ and 
$\bar{\lambda}$, but the term denoted by $B\partial_x\eta_0$ is the same). 
However, at this point we refer the reader to the original articles 
\cite{Shifman:1998zy,Goldhaber:2001rp}.

The local central-charge density has been separately calculated for
the susy case in \cite{Shifman:1998zy}
using higher derivative regularization, and also by
using susy to transform the $\gamma\cdot j$ 
anomaly to the sector with the central
charge. 
The explicit result for the local central-charge 
density of \cite{Shifman:1998zy} 
agrees completely with the explicit result for the energy density of 
\cite{Goldhaber:2001rp} (where
also the explicit local energy density for the non-susy case is
obtained). In \cite{Litvintsev:2000is} a
calculation of the integrated central charge can be found
in global mode regularization, with one
cut-off for the Dirac delta function in the canonical equal-time
(anti)commutation relations and another cut-off for the propagators;
it is
argued that these cutoffs should be the same and this indeed yields the
correct result.
In \cite{Goldhaber:2001rp} the anomaly in the local central charge was obtained
by starting from the definition
\be 
\zeta(y)=\int dx\, \delta(x-y) \{\hal \varphi'(x) U(y)+ \hal
\varphi'(y) U(x)\},
\ee
and not setting $x=y$ too soon.
There is a $1/(x-y)$ singularity in the
propagator $\langle \eta'(x) \eta(y)\rangle$,
and expanding $x$ around $y$ in
the remaining terms, one finds a finite $(x-y)/(x-y)$ term which yields the
anomaly.

So, in conclusion, the nonvanishing one-loop result for the energy density
and total mass of
the minimally susy kink as well as the associated nontrivial modification
of the central charge (density) have been established in the various
regularization methods. The specific subtleties of the different methods
are now well understood, and the origin of the anomalous contribution
to the central charge in each method clarified, which in particular
in dimensional regularization and reduction shows most interesting
facets.

\subsubsection*{Acknowledgments}

It is a pleasure to thank the organizers for heroic
achievements: a wonderful conference without glitches,
an old-Russian-style banquet, and sleeping arrangements for all.


\appendix

\section*{Appendix: The DHN program applied to fermions}

The celebrated DHN calculation of the one loop corrections $M^{(1)}$
to the kink mass \cite{Dashen:1974cj} due to bosonic fluctuations 
has been repeated in \cite{Rebhan:1997iv} for the fermionic case, using exactly the 
same steps as DHN did for bosons. We present this calculation here for two 
reasons: (i) to convince skeptics that there is indeed a problem with the fermions 
if one straightforwardly (or better: naively) repeats the same steps, and (ii) 
because there are subtleties with the zero modes which can be clearly 
illustrated by this concrete example. One might anticipate trouble by 
realizing that for supersymmetric boundary conditions (where all non-zero 
bosonic and fermionic modes cancel pairwise) the result for 
$\Delta M^{(1)}$ would 
be equal to only the counter term $\delta M$ which is divergent.
Some physicists still feel uncomfortable with supersymmetry and prefer to 
stick to older ``reliable'' methods. The following explicit calculation
should make it clear that these older methods need updating, for example 
along the lines suggested in the text.

\begin{figure}
\centerline{
\includegraphics[bb=84 620 240 735,width=4cm]{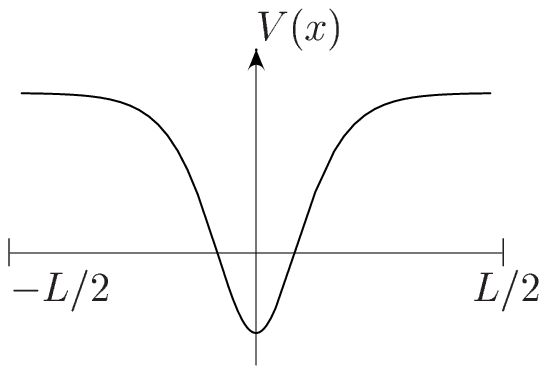}
\includegraphics[bb=84 620 240 735,width=4cm]{v1B.eps}
\includegraphics[bb=84 620 240 735,width=4cm]{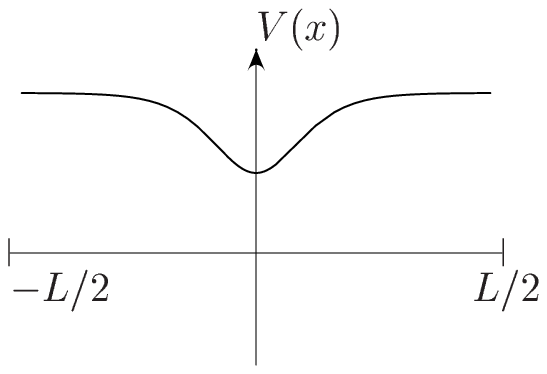}
}
\centerline{\hfil ( $\eta$ ) \hfil \hfil \qquad ( $\psi^+$ ) \hfil\hfil \qquad 
( $\psi^-$ ) \hfil}
\caption{\label{figpots}Potentials for the one bosonic and the
two fermionic fluctuation fields}
\end{figure}

The field equations for the fermions (= for the fermionic fluctuations) read\break
$\not\!\!\partial\psi+U{'}\psi=0$, where 
$U{'}=\frac{\partial}{\partial\varphi}U(\varphi)$. In the representation 
$\gamma^1=\left({1\atop 0}{0\atop{-1}}\right)$ and 
$\gamma^{0}=\left({0\atop 1}{{-1}\atop 0}\right)$ the Dirac equation reads
\begin{equation}
  \label{eq:dirac}
  (\partial_x+U{'})\psi^+-\partial_t\psi^-=0\,,
   \ (-\partial_x+U{'})\psi^-+\partial_t\psi^+=0\,,
   \ \psi=\left({\psi^+}\atop {\psi^-}\right).
\end{equation}
Iterating and setting $\psi^{\pm}=u^{\pm}(x)e^{-i\omega t}$ yields
\begin{eqnarray}
  \label{eq:diracmode}
  &&\left[\partial_x^2+\omega^2-m^2(\frac{3}{2}\tanh^2\frac{mx}{2}
            -\frac{1}{2})\right]u^+ = 0\,, \\
  &&\left[\partial_x^2+\omega^2-m^2(\frac{1}{2}\tanh^2\frac{mx}{2}
            +\frac{1}{2})\right]u^{-} = 0\,.
\end{eqnarray}
The equation for $u^+$ is the same as for the bosonic fluctuations $\eta$
($\varphi=\varphi_K(x)+\eta(x,t)$), and hence before imposing boundary conditions,
the solutions for $u^+$ are the same as for $\eta$. Given a solution for $u^+$ with
$\omega\neq 0$, the Dirac equation gives the corresponding solution for
$u^-$. From the shape of the potentials for the fluctuations
(see Fig.~\ref{figpots}),
it is clear that $\eta$ and $\psi^+$ have a zero mode 
(a normalizable solution of 
the linearized equations for the fluctuations) but $u^-$ has no normalizable
$\omega=0$ solution on $-\infty<x<\infty$.
However, enclosing the system in a box $-L/2\leq x\leq L/2$, also $u^-$ has 
a zero mode.

For the bosonic fluctuations the zero mode 
$\eta_0\sim\cosh^{-2}(\frac{mx}{2})$ 
with strictly $\omega_0=0$ does not satisfy periodic boundary conditions 
because its derivative is odd in $x$, but by slightly increasing the energy 
$\omega$, we can achieve that its derivative vanishes at the boundaries.
Hence, in the bosonic sector there is one almost-zero mode with 
$\omega_0^2\gtrapprox0$.%
\footnote{The other solution for the bosonic fluctuations with 
$\omega=0$, given by 
$m\cosh^{-2}{mx\over 2}\int_0^x dy \cosh^4{my\over 2}$
 does not contribute, even though it is normalizable in the box,
because it is odd in $x$ and does not tend to zero for large $|x|$.
Hence one cannot make it periodic by slightly increasing $\omega^2$.}
In the second-quantized expression for $\eta$ one 
finds then a term $(2\omega_0)^{-1/2}(a_0\eta_0(x)e^{-i\omega_0t}+h.c.)$ 
which appears 
on a par with the genuine non-zero modes, and hence the almost-zero mode should
correspond to one term in the sum over zero point energies, just as DHN
assumed.

\begin{figure}
\centerline{
\includegraphics[bb=84 620 240 735,width=4cm]{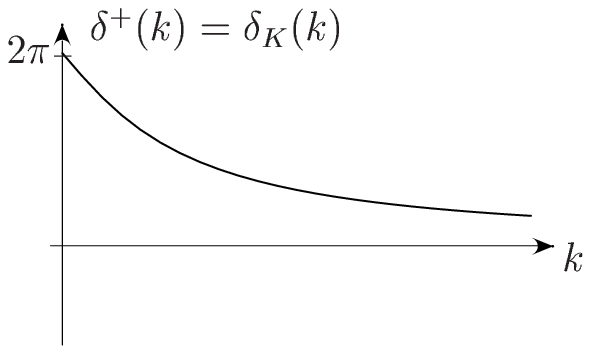}\qquad
\includegraphics[bb=84 620 240 735,width=4cm]{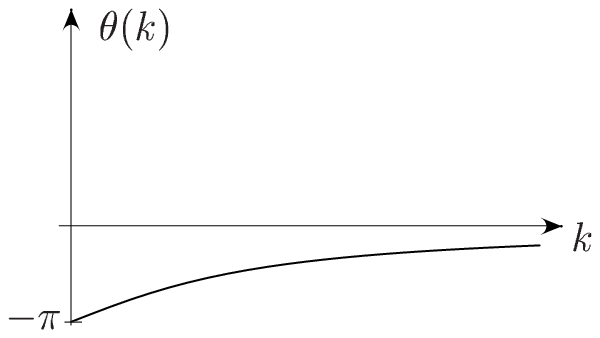}\qquad
\includegraphics[bb=84 620 240 735,width=4cm]{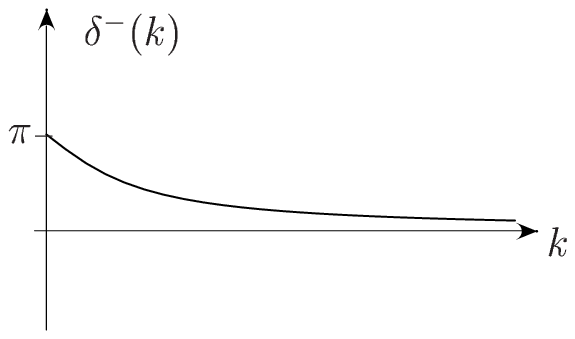}
}
\caption{The phase shift functions $\delta^+\equiv\delta_K$, 
$\theta$, and $\delta^-=\delta^+ + \theta$. \label{figphsh}}
\end{figure}

Imposing even boundary conditions on the fermions%
\footnote{Even boundary conditions are not periodic: the derivatives satisfy 
Robin boundary conditions 
$(\partial_x-m)\psi^+(-L/2)=(\partial_x+m)\psi^+(L/2)$
and $(\partial_x+m)\psi^-(-L/2)=(\partial_x-m)\psi^-(L/2)$ 
because the mass term
$m\tanh\frac{mx}{2}$ switches sign between $-L/2$ and $L/2$.}
\begin{equation}
  \label{eq:fbc}
  \psi^+(-L/2)=\psi^+(L/2) \,,\quad \psi^-(-L/2)=\psi^-(L/2)\,,
\end{equation}
we find the following mode solutions for $\omega\neq0$:
\begin{eqnarray}
  \label{eq:fmods1}
  \psi^+=\cos(kx\pm\hal\delta^+(k))\cos\omega t&\mbox{and}&
   \psi^-=-\sin(kx\pm\hal\delta^-(k))\sin\omega t,\qquad \\
  \label{eq:fmods2}
  \psi^+=-\sin(kx\pm\hal\delta^+(k))\sin\omega t&\mbox{and}&
   \psi^-=\cos(kx\pm\hal\delta^-(k))\cos\omega t,
\end{eqnarray}
with $\pm$ being
a $+$ sign for large positive $x$ and a $-$ sign for large negative $x$.
The Dirac equation is satisfied for $k\geq0$ if
\begin{equation}
  \label{eq:d2f}
  \textstyle\frac{k}{\omega}=\cos\hal\theta(k)\,,\ \frac{m}{\omega}=-\sin\hal\theta(k)\,, \
   \delta^-(k)=\delta^+(k)+\theta(k),
\end{equation}
where $\delta^+(k)=\delta_K(k)$ 
is the phase shift of the bosonic fluctuations.%
\footnote{Its explicit form $\delta_K(k)=-2\arctan(3mk/(m^2-2k^2))$ is not 
needed at this point.} 
The solutions with $k<0$ are obtained from the solutions with $k>0$ by
dropping the two minus signs in (\ref{eq:fmods1}) and (\ref{eq:fmods2}),
but for $k<0$ (\ref{eq:d2f}) becomes $\delta^-(k)=\delta^+(k)+\theta(k)+\pi$
and thus the solutions with $k<0$ are the same as with $k>0$.
The cosines satisfy the boundary conditions, but the sines must vanish 
at the boundaries. This yields two sets of quantization conditions on $k\geq0$:
\begin{equation}
  \label{eq:kqu}
  k_n^+L+\delta^+(k_n^+)=2\pi n^+,\ k_n^-L+\delta^-(k_n^-)=2\pi n^-.
\end{equation}
Given the shape of the phase shifts (see Fig.~\ref{figphsh})
it is clear that $n^-=1,2,3,\dots$, but $n^+=2,3,4,\dots$, 
because the solution with 
$n^+=1$ (yielding $k^+=0$) does not satisfy the boundary conditions.%
\footnote{The solution with $k^+=0$ reads $u^+\sim 3\tanh^2\frac{mx}{2}-1$,
and is even, but $u^-=\frac{i}{\omega}(\partial_x+U{'})u^+$ is odd and does
not vanish for large $|x|$.}

\begin{figure}
\centerline{
\includegraphics[bb=84 620 240 735,width=4cm]{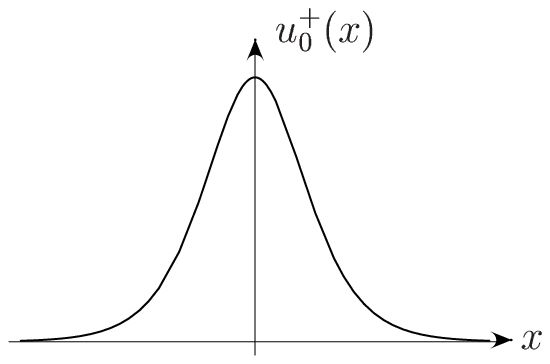}\qquad
\includegraphics[bb=84 620 240 735,width=4cm]{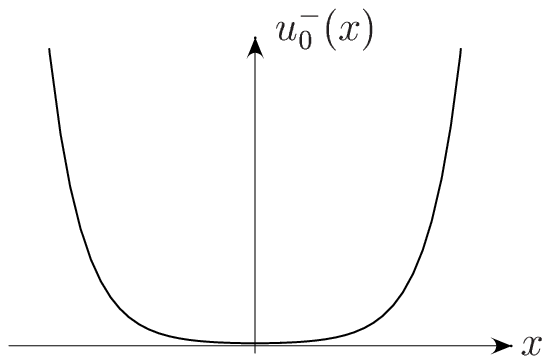}
}
\caption{The fermionic zero modes
 $u^+_0(x)=1/\cosh^2(mx/2), u_0^-=0$ (concentrated at the kink)
and $u_0^+=0, u_0^-=\cosh^2(mx/2)$ (concentrated at the boundary, if any).
\label{figfzm}}
\end{figure}

We now turn to a closer study of the fermionic zero modes. They are both even, 
one concentrated near the kink and the other near the boundaries
(see Fig.~\ref{figfzm}). Zero modes often correspond to symmetries
of the classical action, but the fermionic zero mode which is concentrated
on the boundary is an example of a zero mode which does not correspond
to a symmetry, as one might expect since it ceases to be a normalizable
zero mode on the infinite interval.

The mode expansions of the fermions read
{\setlength\arraycolsep{2pt}
\begin{eqnarray}
  \label{eq:fmodes}
  \psi^+&=&\frac{1}{\sqrt{L}}{\sum_{n>0}}
   \left(c_n\frac{u_n^+(x)}{\sqrt{2}}e^{-i\omega_nt}+
      c_n^\dagger\frac{u_n^+(x)}{\sqrt{2}}e^{i\omega_nt}\right)
             +\frac{c_0u_0^+(x)}{\sqrt{L}}\,,\\
   \psi^-&=&\frac{1}{\sqrt{L}}{\sum_{n>0}}
   \left(c_n\frac{u_n^-(x)}{\sqrt{2}}e^{-i\omega_nt}+
      c_n^\dagger\frac{u_n^-(x)}{\sqrt{2}}e^{i\omega_nt}\right)
             +\frac{d_0u_0^-(x)}{\sqrt{L}}\,,
\end{eqnarray}}%
where the sum over $n$ runs over both sets in (\ref{eq:kqu}), and
where $u_n^\pm(x)$ are normalized to unit-strength plane waves 
$\exp i(k_nx\pm\hal\delta^{\pm}(k_n))$ for large $\pm|x|$. 
Imposing the equal-time 
canonical anti-commutation relations
$\{\psi^\pm(x,t),\psi^\pm(y,t)\}=\delta(x-y)$ 
one finds
\begin{equation}
  \label{eq:acom}
  \frac{1}{L}{\sum_{n>0}}\{c_n,c_n^\dagger\}u_n^+(x)u_n^+(y)+
     \frac{1}{L}\{c_0,c_0\}u_0^+(x)u_0^+(y)=\delta(x-y),
\end{equation}
and a similar relation for $\psi^-$. To determine the value of the mode 
anti-commutators, we need a completeness relation for the mode functions 
$u_n^+(x)$, $u_0^+(x)$. We go back to the second-order differential equation for 
$\psi^+$, and imposing a second boundary condition which follows from the 
Dirac equation
\begin{equation}
  \label{eq:bcf2}
  \psi^+(-L/2)=\psi^+(L/2)\,,\ (\partial_x-m)\psi^+(-L/2)=(\partial_x+m)\psi^+(L/2)\,,
\end{equation}
we obtain a bona-fide selfadjoint elliptic differential operator 
(with bosonic mode operators $a_n$ and $a_n^\dagger$), whose spectrum consists 
of $u_n^+(x)$, $u_0^+(x)$. This proves the completeness relation
\begin{equation}
  \label{eq:compf}
  {\sum_{n>0}}\frac{1}{L}u_n^+(x)u_n^+(y)+\frac{1}{L}u_0^+(x)u_0^+(y)=
    \delta(x-y).
\end{equation}
Comparing with (\ref{eq:acom}) we deduce
\begin{equation}
  \label{eq:acum}
  \{c_m,c_n^\dagger\}=\delta_{m,n}\, ,\ \{c_0,c_0\}=1.
\end{equation}

The Hamiltonian density for fermions
\begin{equation}
  \label{eq:hdensf}
  {\cal{H}}=\frac{i}{2}\psi^T\gamma^0(\gamma^1\partial_x+U')\psi=
    \frac{i}{2}(\psi^+\partial_t\psi^++\psi^-\partial_t\psi^-)
\end{equation}
yields the expected negative contribution to the zero-point energy, 
$\left<H\right>=\break-\hal\hbar{\sum_{n>0}}\omega_n$, and in the
density $\left<\psi^+(x)\psi^+(x)\right>$ the zero mode contributes
a term $\hal\frac{1}{L}u_0^+(x)u_0^+(x)$. Due to this factor $\hal$, the two 
zero modes of the fermions in a box with even boundary conditions contribute 
one term to the sum over zero-point energies, just as 
for the bosonic case, and just as implicitly assumed by DHN.%
\footnote{One can give a better argument. 
Deforming the potential for the fermions
slightly the two zero modes become one almost-zero mode, on a par with the 
other genuine nonzero modes.}

We can now compute $\Delta M^{(1)}$. 
The fermionic modes cancel half of the bosonic modes 
for $n\geq2$, but the bosonic $n=1$ mode is left. The bound states and the 
zero modes cancel between bosons and fermions. This yields
\begin{eqnarray}
  \label{eq:M1}
  \Delta M^{(1)}&=&\hal{\sum_{n\geq 1}}\omega(k_n^+)-
    \hal{\sum_{n\geq 2}}\omega(k_n^-)+\delta M \nonumber\\
    &=&\hal m+\hal{\sum_{n\geq2}}\frac{\partial\omega(k_n^+)}{\partial k}
       \frac{\theta(k_n)}{L}+\delta M \nonumber\\
    &=&\hal m+\hal\int_0^\infty\frac{dk}{2\pi}\frac{\partial\omega}{\partial k}
        \theta(k)+\delta M=\frac{\pi-2}{4\pi}m,
\end{eqnarray}
where we used that $k_n^+=k_n^-+\theta(k_n^-)/L$. This is the correct answer to
an incorrect question, because this value for $M^{(1)}$ contains spurious 
boundary energy. In the text it is discussed how to separate off the
spurious boundary energy, and the correct result is
\begin{equation}
  \label{eq:res}
  M^{(1)}=-\frac{m}{2\pi}\,.
\end{equation}
If one repeats the same calculations for the sine-Gordon system, one can compare 
with the exact result obtained from the Yang-Baxter equation, and finds indeed
that these two results agree after removing the boundary energy.

\end{document}